\documentclass{icrc29}
\usepackage{graphicx,amssymb,amsmath,times}
\begin{document}
\title[Probing for Leptonic Signatures from GRB030329 with AMANDA-II]{Probing for Leptonic Signatures from GRB030329 with AMANDA-II\footnote{\emph{Adapted from a contribution to the Proceedings of the 29$^{\text{th}}$ International Cosmic Ray Conference in Pune, India (August 2005)}.}}

\author[M. Stamatikos et al.] {Michael Stamatikos$^a$ for the IceCube Collaboration$^b$, Jenny Kurtzweil$^a$ and Melanie J. Clarke$^a$
        \newauthor
            \\
        (a) Department of Physics, University of Wisconsin, Madison, 1150 University Avenue, Madison, WI 53706, USA \\
        (b) For a full author list see astro-ph/0509330
        }
\presenter{Presenter: M. Stamatikos (michael.stamatikos@icecube.wisc.edu), \
usa-stamatikos-M-abs1-og24-oral}

\maketitle

\begin{abstract}

The discovery of high-energy (TeV-PeV) neutrinos from gamma-ray bursts (GRBs) would shed light on their intrinsic microphysics by confirming hadronic acceleration in the relativistic jet; possibly revealing an acceleration mechanism for the highest energy cosmic rays. We describe an analysis featuring three models based upon confronting the fireball phenomenology with ground-based and satellite observations of GRB030329, which triggered the High Energy Transient Explorer (HETE-II). Contrary to previous diffuse searches, the expected \emph{discrete} muon neutrino energy spectra for models 1 and 2, based upon an isotropic and beamed emission geometry, respectively, are directly derived from the fireball description of the prompt $\gamma$-ray photon energy spectrum, whose spectral fit parameters are characterized by the Band function, and the spectroscopically observed redshift, based upon the associated optical transient (OT) afterglow. For comparison, we also consider a model (3) based upon \emph{averaged} burst parameters and isotropic emission. Strict spatial and temporal constraints (based upon electromagnetic observations), in conjunction with a single, robust selection criterion (optimized for discovery) have been leveraged to realize a nearly background-free search, with nominal signal loss, using archived data from the Antarctic Muon and Neutrino Detector Array (AMANDA-II). Our preliminary results are consistent with a null signal detection, with a peak muon neutrino effective area of $\sim80$ m$^2$ at $\sim 2$ PeV and a flux upper limit of $\sim0.150\:\text{GeV}/\text{cm$^{2}/$s}$ for model 1. Predictions for IceCube, AMANDA's kilometer scale successor, are compared with those found in the literature. Implications for correlative searches are discussed.

\end{abstract}

\section{Introduction}

Neutrino astronomy may provide us with a new glimpse at the internal processes of gamma-ray bursts (GRBs). The Antarctic Muon and Neutrino Detector Array (AMANDA), which has been calibrated upon atmospheric neutrinos, has demonstrated the viability of high energy neutrino astronomy by using the ice at the geographic South Pole as a Cherenkov medium. Canonical fireball phenomenology \cite{Piran:2004}, in the context of hadronic acceleration, predicts correlated MeV-EeV neutrinos from GRBs via various hadronic interactions \cite{Stamatikos:2004}. Ideal for detection are TeV-PeV neutrinos in coincidence with prompt $\gamma$-ray emission, resulting in a nearly background-free search. A detailed description of the modeling techniques and an ongoing analysis featuring correlated neutrino searches of individual GRBs from the Burst and Transient Source Experiment (BATSE) are described elsewhere \cite{Stamatikos:2005,Stamatikos:2004}. Here, we report on a complementary search for correlated leptonic ($\nu_{\mu},\bar{\nu}_{\mu}$) emission, using models based upon the unique \emph{discrete} electromagnetic characteristics and emission geometry of GRB030329, gleaned directly from satellite and ground-based observations. This represents a novel departure from searches~\cite{Hardtke:2003,Kuehn:2005,Hughey:2005} based upon a diffuse formulation \cite{WB:1999}, which utilize averaged burst parameters.

\section{GRB030329: Electromagnetic Emission \& The GRB-Neutrino Connection}

On March 29, 2003, at $11^{\text{h}}37^{\text{m}}14.^{\text{s}}67$ (UTC), HETE-II was triggered by GRB030329 (H2652), a watershed transient which confirmed the connection between a core collapse Type Ic supernova and long duration GRB.  Electromagnetic investigations of the prompt $\gamma$-ray and multi-wavelength afterglow emissions associated with GRB030329 abound in the literature (see table~\ref{GRB030329_electromagnetic_properties}), making it a perfect specimen for study. Via forward folding deconvolution, the photon energy spectrum was fit to the \emph{Band function} \cite{Band:1993}, an empirically derived power law with smooth transition. For spectral indices $\alpha > -2$ and $\beta < -2$, the characteristic peak energy is defined as $E_{p}=[(2+\alpha)\epsilon_{\gamma}^{b}](\alpha-\beta)^{-1}$, where $\epsilon_{\gamma}^{b}$ is known as the photon break energy \cite{Band:1993}. Hence, using standard error propagation, we find that $\epsilon_{\gamma}^{b}=115.6\pm9.9$ keV. Doppler spectroscopy of the OT afterglow revealed a redshift measurement, which, under an assumed $\Lambda_{\text{CDM}}$ cosmology,\footnote{$\Lambda_{\text{CDM}}$ cosmology: H$_{\text{o}}$ = 72 $\pm$ 5 km/Mpc/s, $\Omega_{\text{m}}$ = 0.29 $\pm$ 0.07, $\Omega_{\Lambda}$ = 0.73 $\pm$ 0.07 [Spergel et al., ApJS 148, 175-194 (2003)].} placed GRB030329 at a luminosity distance of $2.44^{+0.20}_{-0.18}\times10^{27}$ cm. Coupled with the peak energy flux, this implies an intrinsic peak isotropic luminosity of $L_{\gamma}^{iso}\approx 5.24^{+0.86}_{-0.77}\times10^{50}\:\text{ergs}/\text{s}$ in the 30-400 keV energy band pass. Evidence for anisotropic emission, in the form of a two component break in the afterglow spectrum, was revealed by radio calorimetry and is consistent with collimated prompt emission within a jet of opening half angle $\theta_{jet}$. This requires a beaming fraction correction, which reduces the intrinsic peak luminosity to $L_{\gamma}^{jet}=L_{\gamma}^{iso}(1-\cos\theta_{jet})\approx 1.99^{+0.33}_{-0.29}\times10^{48}\:\text{ergs}/\text{s}$. Extended calorimetry provided an estimate for the fractions of shock energy imparted to the electrons ($\epsilon_{e}$) and magnetic field ($\epsilon_{B}$). Table~\ref{GRB030329_electromagnetic_properties} summarizes the observed electromagnetic properties used in this analysis.

\begin{table} [h]
\begin{center}
\begin{minipage}[c]{1.00\textwidth}
\caption [l] {Electromagnetic Properties of GRB030329: Prompt $\gamma$-ray and Multi-wavelength Afterglow Emission}
\smallskip
\begin{tabular}{|c|c|c|}
\hline
\small{\textbf{Parameter(s)}} & \small{\textbf{Value(s)}} & \small{\textbf{Reference}}\\
\hline
\scriptsize{Positional Localization $\left(\alpha_{\text{J2000}},\delta_{\text{J2000}},\sigma_{R}\right)$} &  \scriptsize{$161.2081646^{\circ},21.5215106^{\circ},3.0\times10^{-7}\:^{\circ}$} & \scriptsize{Taylor et al., GCN Report 2129} \\
\scriptsize{Trigger (T) \& Duration (T$_{90}$) [30-400 keV]} & \scriptsize{$41,834.67\:\text{SOD},22.8\pm0.5\:\text{s}$} & \scriptsize{Vanderspek et al., ApJ 617: 1251-1257 (2004)} \\
\scriptsize{Energy Fluence ($F_{\gamma}$) [2-400 keV]} & \scriptsize{$1.630^{+0.014}_{-0.013}\times10^{-4}\:\text{ergs}/\text{cm$^2$}$} & \scriptsize{Sakamoto et al., astro-ph/0409128} \\
\scriptsize{Band Parameters $\left(\alpha,\:\beta,\:E_{p}\right)$ [2-400 keV]} & \scriptsize{$-1.32\pm0.02$,$-2.44\pm0.08$, $70.2\pm2.3$} & \scriptsize{Vanderspek et al., ApJ 617: 1251-1257 (2004)} \\
\scriptsize{Peak Energy Flux ($\Phi_{\gamma}^{\text{Peak}}$) [30-400 keV]} & \scriptsize{$\sim7\times10^{-6}\:\text{ergs}/\text{cm$^{2}$/s}$} & \scriptsize{Vanderspek et al., GCN Report 1997} \\
\scriptsize{Spectroscopic Redshift (z)} & \scriptsize{$0.168541\pm0.000004$} & \scriptsize{Bloom et al., GCN Report 2212} \\
\scriptsize{Jet Opening Half Angle ($\theta_{jet}$)} & \scriptsize{$\sim5^{\circ}\approx0.09$ rad} & \scriptsize{Berger et al., Nature 426, 154-157 (2003)} \\
\scriptsize{Electron \& Magnetic Field Energy Fractions } & \scriptsize{$\epsilon_{e}\approx0.19$, $\epsilon_{B}\approx0.042$} & \scriptsize{Frail et al., ApJ 619, 994-998 (2005)} \\
\hline
\end{tabular}
\label{GRB030329_electromagnetic_properties}
\end{minipage}
\end{center}
\end{table}

The generic mechanism responsible for the super-Eddington luminosities associated with GRBs is the dissipation, via shocks, of highly relativistic kinetic energy, acquired by electrons and positrons Fermi accelerated in an optically thick, relativistically expanding plasma, commonly referred to as a \emph{fireball}. The acceleration of electrons in the intense magnetic field of the fireball leads to the emission of prompt non-thermal $\gamma$-rays via synchrotron radiation. The temporal variability ($t_{v}\sim 10$ ms) in the light curves imply compact sources. In order to ensure a transparent optical depth to photons of energy $\epsilon_{\gamma}^{max}\approx100\:\text{MeV}$, a minimum bulk Lorentz boost factor ($\Gamma$) was assigned (see equation~\ref{Gamma}). Hadronic acceleration within the ambient photon field produces TeV-PeV leptons, via the following photomeson interaction:

\begin{equation}
p+\gamma \rightarrow \Delta ^{+} \rightarrow \pi ^{+} + [n] \rightarrow \mu^{+}+\nu_{\mu} \rightarrow e^{+} + \nu_e + \bar{\nu}_{\mu}+\nu_{\mu}
\label{photomeson}
\end{equation}

Hence, these neutrinos are expected to be spatially and temporally correlated with prompt $\gamma$-ray emission. The parameterization of the neutrino energy spectra $(dN_{\nu_{\mu}}/d\epsilon_{\nu_{\mu}}\equiv\Phi_{\nu_{\mu}})$ traces the prompt photon energy spectra, as illustrated in figure~\ref{neutrino_spectrum_and_response}, and is defined as follows \cite{Stamatikos:2005,Guetta:2004}:

\begin{subequations}

\begin{equation}
\epsilon_{\nu_{\mu}}^{2}\Phi_{\nu_{\mu}}\approx A_{\nu_{\mu}}\times
\begin{cases}
\left(\frac{\epsilon_{\nu_{\mu}}}{\epsilon_{\nu}^{b}}\right)^{-\beta-1} & \epsilon_{\nu_{\mu}}<\epsilon_{\nu}^{b} \\
\left(\frac{\epsilon_{\nu_{\mu}}}{\epsilon_{\nu}^{b}}\right)^{-\alpha-1} & \epsilon_{\nu}^{b}<\epsilon_{\nu_{\mu}}<\epsilon_{\pi}^{b} \\
\left(\frac{\epsilon_{\nu_{\mu}}}{\epsilon_{\nu}^{b}}\right)^{-\alpha-1}\left(\frac{\epsilon_{\nu_{\mu}}} {\epsilon_{\pi}^{b}}\right)^{-2} & \epsilon_{\nu_{\mu}}>\epsilon_{\pi}^{b}
\end{cases}
\label{neutrino_energy_spectrum}
\end{equation}

\begin{equation}
A_{\nu_{\mu}}\approx\frac{F_{\gamma} f_{\pi}}{8\epsilon_{e}\ln(10)T_{90}}\approx9.86\times10^{-4}\:\text{GeV}/\text{cm$^{2}/$s}
\label{neutrino_normalization}
\end{equation}

\begin{equation}
f_{\pi}\simeq 0.2\times \frac{L_{\gamma,52}}{\Gamma_{2.5}^{4}t_{v,-2}\epsilon_{\gamma, MeV}^{b}(1+z)}\approx0.77
\label{proton_efficiency}
\end{equation}

\begin{equation}
\Gamma\gtrsim276\left[L_{\gamma,52}t_{v,-2}^{-1}\epsilon_{\gamma,MeV}^{max}(1+z)\right]^{\frac{1}{6}}\approx178
\label{Gamma}
\end{equation}

\begin{equation}
\epsilon_{\nu}^{b}\approx\left[\dfrac{7\times10^{5}}{(1+z)^{2}} \dfrac{\Gamma_{2.5}^{2}}{\epsilon_{\gamma,\text{MeV}}^{b}}\right]\:\text{GeV}\approx1.404951\times10^{6}\:\text{GeV}
\label{neutrino_break_energy}
\end{equation}

\begin{equation}
\epsilon_{\pi}^{b}\approx\left[\dfrac{10^{8}}{(1+z)}\epsilon_{e}^{\frac{1}{2}}\epsilon_{B}^{-\frac{1}{2}} (L_{\gamma,52})^{-\frac{1}{2}}\Gamma_{2.5}^{4}t_{v,-2}\right]\:\text{GeV}\approx7.9832941\times10^{7}\:\text{GeV}
\label{synchrotron_break_energy}
\end{equation}

\end{subequations}

The values in equations \ref{neutrino_normalization}-\ref{synchrotron_break_energy} are given for model 1, where $L_{\gamma}\equiv L_{\gamma,52}\cdot10^{52}\:\text{ergs/s}$, $\Gamma\equiv\Gamma_{2.5}\cdot10^{2.5}$, $t_{v}\equiv t_{v,-2}\cdot 10\:\text{ms}$, $\epsilon_{\gamma}^{b}\equiv\epsilon_{\gamma,\text{MeV}}^{b}\cdot1\:\text{MeV}$, and $\epsilon_{\gamma}\equiv\epsilon_{\gamma,MeV}^{max}\cdot100\:\text{MeV}$. Note the explicit dependence on discrete $\gamma$-ray photon observables.

\section{Neutrino Astronomy with AMANDA-II: Analysis, Results \& Discussion}

\begin{figure}[h]
\begin{center}
\includegraphics*[width=1.0\textwidth,angle=0,clip]{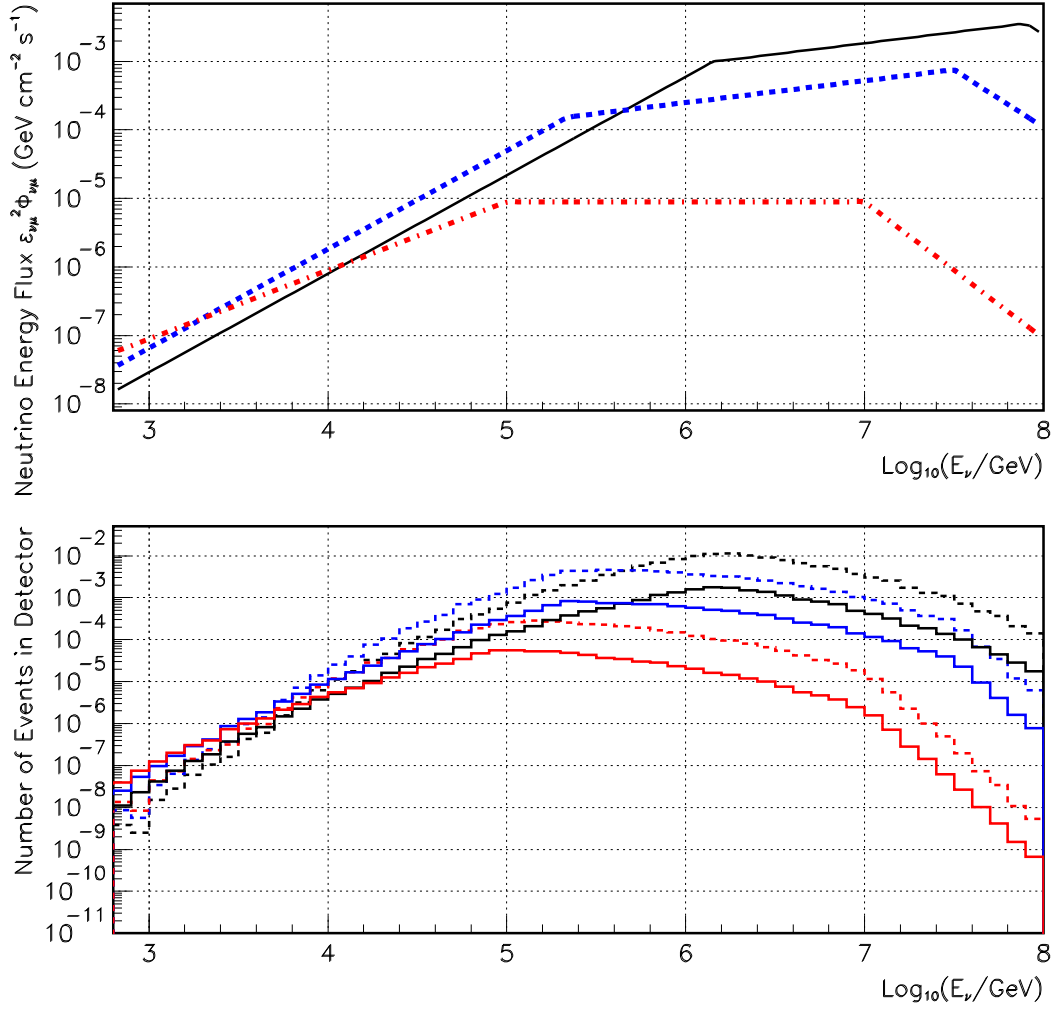}
\caption{\label {neutrino_spectrum_and_response} \emph{Upper panel} - Prompt neutrino energy flux for models 1 (solid black), 2 (dashed blue) and 3 (dot-dashed red), based upon equation~\ref{neutrino_energy_spectrum}, for GRB030329. \emph{Lower panel} - Detector response for models 1 (black), 2 (blue) and 3 (red) for AMANDA-II (solid) and IceCube (dashed). Note how the variance in the neutrino spectra (upper panel), due to the variance of electromagnetic parameters used in equation~\ref{neutrino_energy_spectrum}, directly translate into the variance in detector response (lower panel), which exceed an order of magnitude, as manifested in observables such as the expected mean neutrino energy and rate. The effects of neutrino flavor oscillations have been included.}
\end{center}
\end{figure}

AMANDA-II is comprised of 677 optical modules buried at depths between 1500-2000 m. The background consists of cosmic ray induced \emph{down-going} atmospheric muons, detected at a rate of $\sim100$ Hz, with a perturbation of atmospheric neutrinos, detected at a rate of $\sim10^{-4}$ Hz. The astrophysical neutrino signal, detected via charged current interactions such as: $\nu_{\mu}+N\rightarrow\mu^{\pm}+X$, is isolated by utilizing topologically identified \emph{up-going} muon events, which are reconstructed by a maximum likelihood method. On-source, off-time data were used to estimate the stability of the background rate in order to maintain blindness, facilitating an unbiased analysis. After data filtering (see \cite{Ackermann:2005}), the total off-time background, excluding a 10 minute blinded window centered upon the trigger time, was consistent with a Gaussian fit and accrued $24,972\pm158$ events over a 57,328.04 second interval, resulting in a rate of $0.436\pm0.003$ Hz. Based upon a visual inspection of the light curve, a conservative search window of 40 seconds (beginning at $T$) was chosen. Hence, $17.44\pm0.12$ background events ($n_{b}$) were expected on-time in AMANDA-II prior to any quality selection. Signal neutrino spectra were simulated for three models (see figure~\ref{neutrino_spectrum_and_response}) by propagating a total of $\sim440,000$ muon neutrinos ($\nu_{\mu}+\bar{\nu}_{\mu}$) from an error box in the sky defined by the spatial localization of the radio afterglow (see table~\ref{GRB030329_electromagnetic_properties}). Event quality selection was optimized for the best limit setting and discovery potential by minimizing the model rejection factor (MRF) \cite{Hill:2003} and the model discovery factor (MDF) \cite{Hill:2005}, respectively. Although multiple observables were investigated, a single criterion emerged, based upon the maximum size of the search bin radius ($\Psi$), i.e. the space angle between the reconstructed muon trajectory and the GRB's position.

\begin{figure}[h]
\begin{center}
\includegraphics*[width=1.0\textwidth,angle=0,clip]{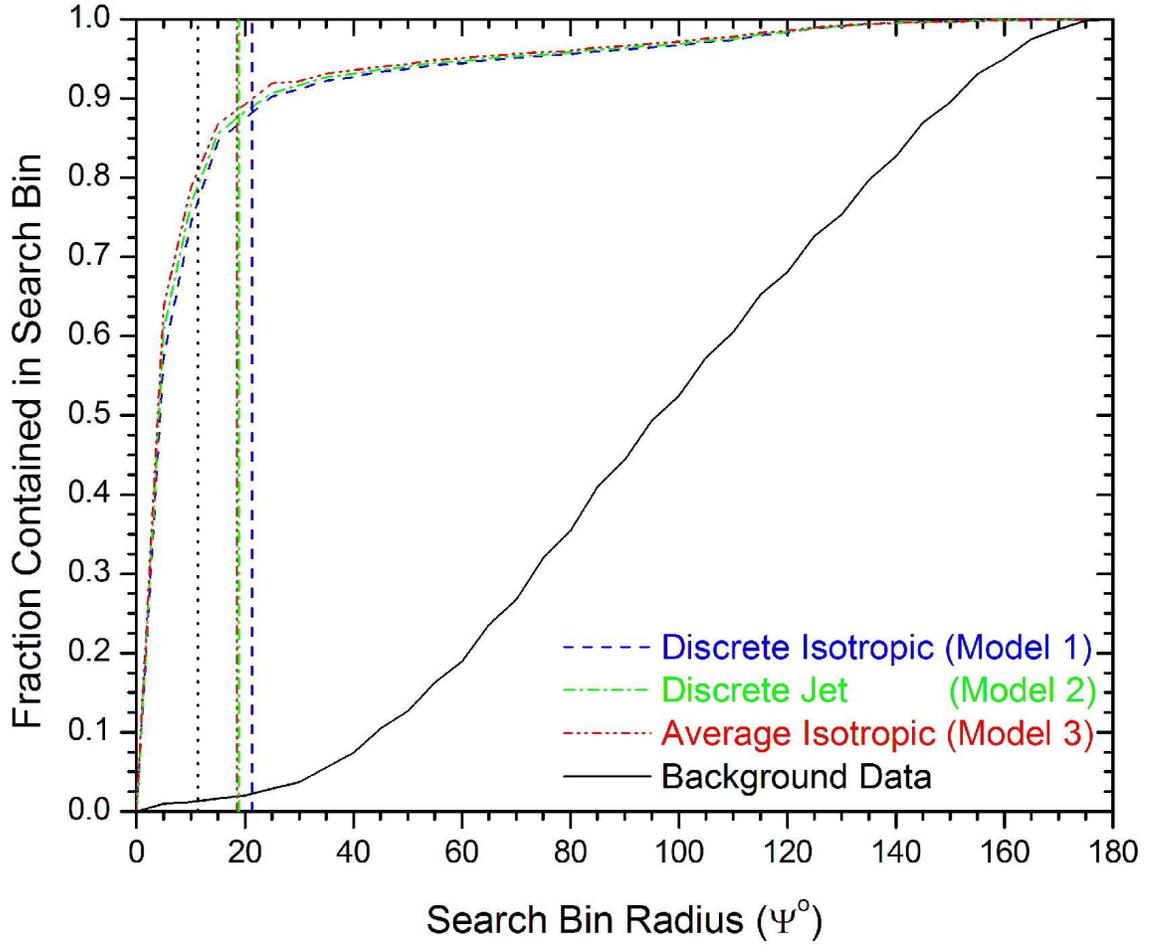}
\caption{\label {frac_v_psi} AMANDA-II signal efficiency/background rejection for GRB030329 models 1-3 using MRF (vertical dashed blue, dashed-dot green and dashed-dot-dot red) and MDF (vertical black dotted) optimizations. Both MRF and MDF selections reject $\sim99\%$ of the background while retaining $\sim86\%$ and $\sim77\%$ of the signal, respectively.}
\end{center}
\end{figure}

\begin{figure}[h]
\begin{center}
\includegraphics*[width=1.0\textwidth,angle=0,clip]{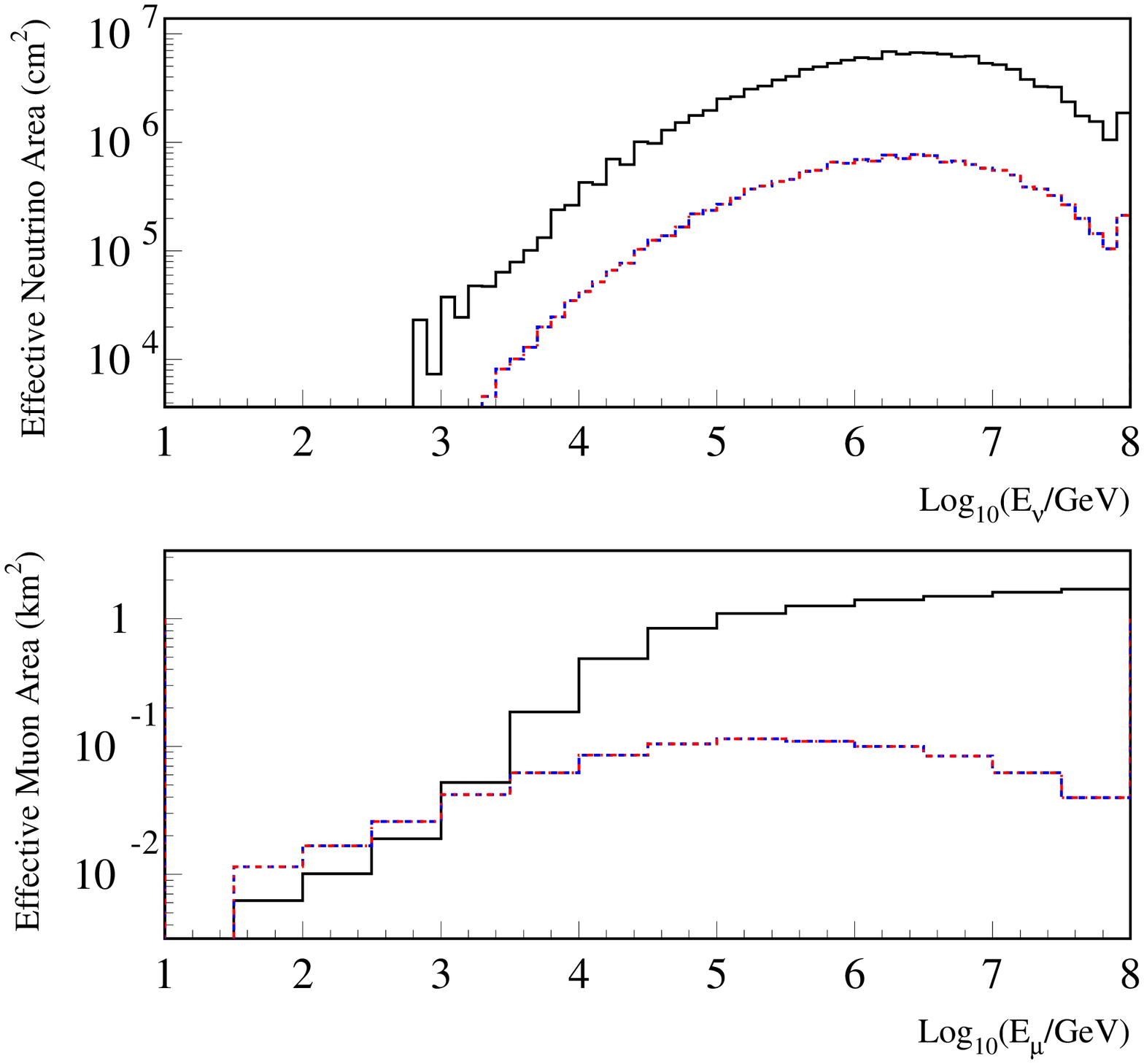}
\caption{\label {neutrino_effective_area} \emph{Upper panel} - The effective muon neutrino area. A peak of $\sim80\:\text{m}^{2}$ and $\sim700\:\text{m}^{2}$ occurs at $\sim2$ PeV for AMANDA-II and IceCube, respectively. \emph{Lower panel} - The effective muon area for energy at closest approach to the detector. At $\sim200$ TeV, the effective muon area for IceCube, $\sim1\:\text{km}^{2}$, exceeds that of AMANDA-II, $\sim100,000\:\text{m}^{2}$, by an order of magnitude. For both panels, MDF optimized AMANDA-II results for models 1 (dashed black), 2 (dashed blue) and 3 (dot-dashed red) and predicted IceCube (solid black) curves are illustrated for GRB030329 with $\delta_{\text{J2000}}\approx22^{\circ}$.}
\end{center}
\end{figure}

Our search, optimized for $5\sigma$ discovery (requiring 4 events within $\Psi\leq11.3^{\circ}$), is consistent with a null signal. Upper limits, summarized with our results in table~\ref{results}, do not constrain the models tested using AMANDA-II. Effective neutrino and muon areas are given in figure~\ref{neutrino_effective_area}. The number of expected signal events in IceCube ($N_{s}$) for model 1 is consistent with \cite{Razzaque:2004}, when neutrino oscillations are considered. For model 3, our results for $N_{s}$ are in agreement with \cite{Guetta:2004} and \cite{Ahrens:2004}, when one adjusts for the assumptions of \cite{WB:1999}. Selection based upon $\Psi$ was robust across the models tested in AMANDA-II, as illustrated in figure~\ref{frac_v_psi}. However, the MRF/MDF and hence the limit setting/discovery potential was strongly model dependent, varying by over an order of magnitude for models 1 and 3. Furthermore, using the same theoretical framework, the response of AMANDA-II and IceCube to spectra based upon discrete and average parameters are discrepant in mean neutrino energy and event rate by over an order of magnitude as illustrated in figure~\ref{neutrino_spectrum_and_response} and table~\ref{results}. Such variance in detector response unequivocally demonstrates the value of a discrete modeling approach when making correlative neutrino observations of individual GRBs, especially in the context of inferred astrophysical constraints, in agreement with \cite{Guetta:2004}.

\begin{table} [h]
\begin{center}
\begin{minipage}[c]{1.00\textwidth}
\caption [c] {Summary of Results for GRB030329. Primes indicate value after selection for IceCube (N$_{x}$) and AMANDA-II (n$_{x}$). Superscripts indicate A=MRF and B=MDF optimization. Average upper limit (sensitivity) and flux upper limit values are given for AMANDA-II. The effects of neutrino flavor oscillations have been included.}
\smallskip
\begin{tabular}{|ccccccc|}
\hline
\scriptsize{$\begin{array}{c}
\text{Neutrino}\\
\text{Flux}\\
\text{Model}\\
\text{$\Phi_{\nu_{\mu}}$}\\
\end{array}$} & \scriptsize{$\begin{array}{c}
\text{Search}\\
\text{Bin}\\
\text{Radius}\\
\text{$\Psi^{A},\Psi^{B}\:^{\circ}$}\\
\end{array}$} & \scriptsize{$\begin{array}{c}
\text{Background}\\
\text{Expected in}\\
\text{AMANDA-II}\\
\text{$n_{b},n_{b}^{A\prime},n_{b}^{B\prime}$}\\
\end{array}$} & \scriptsize{$\begin{array}{c}
\text{Expected Number}\\
\text{of Neutrino}\\
\text{(Signal) Events}\\
N_{s},\:n_{s},\:n_{s}^{B\prime}\\
\end{array}$} & \scriptsize{$\begin{array}{c}
\text{Observed}\\
\text{Events in}\\
\text{AMANDA-II}\\
\text{$n_{obs},n_{obs}^{B\prime}$}\\
\end{array}$} & \scriptsize{$\begin{array}{c}
\text{Optimization}\\
\text{Method for}\\
\text{AMANDA-II}\\
\text{MRF, MDF}\\
\end{array}$} & \scriptsize{$\begin{array}{cc}
\text{Average} & \text{Flux} \\
\text{Upper limit$^{B}$} & \text{Upper} \\
\text{[Sensitivity]} & \text{Limit} \\
\text{$\frac{\text{GeV}}{\text{cm$^{2}\cdot$s}}$} & \text{$\frac{\text{GeV}}{\text{cm$^{2}\cdot$s}}$} \\
\end{array}$} \\
\hline
\scriptsize{1} & \scriptsize{21.3, 11.3} & \scriptsize{17.44, 0.23, 0.06} & \scriptsize{0.1308, 0.0202, 0.0156} & \scriptsize{15, 0} & \scriptsize{152, 424} & \scriptsize{$\begin{array}{cccc}
\text{0.157} & & & \text{0.150}\\
\end{array}$} \\
\scriptsize{2} & \scriptsize{18.8, 11.3} & \scriptsize{17.44, 0.17, 0.06} & \scriptsize{0.0691, 0.0116, 0.0092} & \scriptsize{15, 0} & \scriptsize{256, 716} & \scriptsize{$\begin{array}{cccc}
\text{0.041} & & & \text{0.039}\\
\end{array}$} \\
\scriptsize{3} & \scriptsize{18.5, 11.3} & \scriptsize{17.44, 0.17, 0.06} & \scriptsize{0.0038, 0.0008, 0.0006} & \scriptsize{15, 0} & \scriptsize{3864, 10794} & \scriptsize{$\begin{array}{cccc}
\text{0.036} & & & \text{0.035}\\
\end{array}$}\\
\hline
\end{tabular}
\label{results}
\end{minipage}
\end{center}
\end{table}

\section{Acknowledgements}
M. Stamatikos would like to thank K. Hurley and D.L. Band for valuable discussions regarding the electromagnetic properties of GRB030329 and prompt photon energy spectral fits, respectively. Additionally, funding support agencies have been cited in astro-ph/0509330.


\begin{thebibliography}{99}

\bibitem{Piran:2004}
T. Piran, Rev. Modern Phys., 76, 1143-1210 (2005)

\bibitem{Stamatikos:2004}
M. Stamatikos et al., AIP Conf. Proc. 727, 146-149 (2004)

\bibitem{Stamatikos:2005}
M. Stamatikos, D.L. Band, D. Hooper and F. Halzen, in preparation.

\bibitem{Hardtke:2003}
R. Hardtke, K. Kuehn, and M. Stamatikos et al. Proceedings of the 28th ICRC, 2717-2720 (2003)

\bibitem{Kuehn:2005}
K. Kuehn et al., these proceedings.

\bibitem{Hughey:2005}
B. Hughey and I. Taboada et al., these proceedings.

\bibitem{WB:1999}
E. Waxman \& J. Bahcall, Phys. Rev. D 59, 023002 (1999)

\bibitem{Band:1993}
D.L. Band et al., ApJ 413, 281-292 (1993)

\bibitem{Guetta:2004}
D. Guetta et al., Astrop. Phys. 20, 429-455 (2004)

\bibitem{Ackermann:2005}
M. Ackermann et al., These proceedings.

\bibitem{Hill:2003}
G.C. Hill \& K. Rawlins Astropart. Phys. 19 393-402 (2003)

\bibitem{Hill:2005}
G.C. Hill, J. Hodges and M. Stamatikos, in preparation.

\bibitem{Razzaque:2004}
S. Razzaque et al., Phys. Rev. D 69, 023001 (2004)

\bibitem{Ahrens:2004}
J. Ahrens et al., Astrop. Phys. 20, 507-532 (2004)

\end{thebibliography}
\end{document}